%Paper: nucl-th/9308005
%From: Avraham Rinat <FNRINAT@WEIZMANN.weizmann.ac.il>
%Date: Wed, 04 Aug 93 13:00:45 +0300
%Date (revised): Sat, 14 Aug 93 22:06:23 +0300

%macropackage=wizzle
\input mimac
\tolerance 1500
\topskip .75truein

\font\twelverm =cmr12

\def\Oscr{{\cal O}}

\def\p3{{d^3p\over {(2\pi)^3}}}
\def\ff{e_0(\vec p)}
\def\to{\rightarrow}
\def\ee{e_0(\vec p+\vec q)}

\def\tt{t^{ons}}
\def\zz{\langle g\rangle}
\def\qs{\lower 5pt\hbox to 23pt{\rightarrowfill}\atop
       {s\rarrow\infty}}
\def\qq{\lower 5pt\hbox to 23pt{\rightarrowfill}\atop
      {q\rarrow\infty}}
\def\q|s|{\lower 5pt\hbox to 23pt{\rightarrowfill}\atop
        {|s|\rarrow\infty}}
\def\ton{\lower 5pt\hbox to 23pt{\rightarrowfill}\atop
       {t\rarrow \tt}}
\def\gon{\lower 5pt\hbox to 23pt{\rightarrowfill}\atop
       {g\rarrow \zz}}

\def\o{\over}
\twelverm
\newdimen\offdimen
\def\offset#1#2{\offdimen #1
   \noindent \hangindent \offdimen
   \hbox to \offdimen{#2\hfil}\ignorespaces}
\parskip 0pt
\rightline{WIS-93/72/Jul PH}
\vskip .6truein

\def\endpage{\par\vfill\eject}
\def\ctrline{\centerline}
\def\({\lbrack}
\def\){\rbrack}

% This is for reference numbers
\par
\def\tindent#1{\indent\llap{#1}\ignorespaces}
\def\refn{\par\hang\tindent}
\input tables
\baselineskip 24pt
\centerline{\bf Inclusive scattering of electrons from nuclear matter}
\centerline {\bf in the GeV range}
\bigskip
\centerline {A.S. Rinat and M.F. Taragin}
\centerline {Weizmann Institute of Science, Rehovot 76100, Israel}

\vskip 1truein\noindent
\baselineskip 21pt

{\bf Abstract:}

We  study Final  State Interaction  effects (FSI)  in inclusive  electron
scattering of  several GeV from  nuclear matter.  We  consider separately
the cases of $'$elastic$'$ and $'$inelastic$'$ knockout of respectively a
nucleon and a  non-nucleonic particle.  For the former  category FSI have
been calculated, using the first cumulant approximation of a relativistic
generalization of  the theory of Gersch  $et\,al$, including intermediate
$NN$  states states  with produced  hadrons.   We also  discuss a  formal
expression for the first cumulant approximation on the basis of the PWIA.
FSI between  an inelasticlly knocked-out  particle and core  nucleons are
neglected.  We present  two scaling analyses of the data,  from which the
above  $'$inelastic$'$ contributions  have been  been removed.   The data
expressed in a  relativistic West variable show scaling of  a much better
quality  than the  same  in  the IA  scaling  variable.  Predictions  for
inclusive cross sections in the two first cumulant versions are presented
and discussed.   We make a critical  comparison of the above  with the IA
approach of Benhar $et\,al$ and their results.

\vskip.8truein\baselineskip 12pt
%\leftline{\hskip38pt PACS:xxxx}

\vfil

\eject\

\baselineskip 24pt

{\bf 1. Introduction}
\par
We address  in the  following total inclusive  scattering from  nuclei of
electrons with 3-momentum transfer $q$  in the range 0.5  $\le q({\rm
GeV}) \le $2.4 and energy  losses $\omega \le 2.0 $GeV $^1$.
\footnote
* {Protracted illness of one of the authors has caused considerable delay
in  the preparation  of this  ms.   It supersedes  a preliminary  version
WIS-91/42/Aug PH, which circulated two  years ago.}
Under those kinematic
conditions one expects relativistic effects to play a role.

With   only  a   few  theories   available  for   relativistic  many-body
systems$^2$, a description of the  above data poses grave problems.  This
is reflected in studies of the total inclusive (TI) scattering, which are
usually  limited to  the  lightest targets  or  to relativistic potential
models $^3$.

Simplifications  occur if  translational  invariance  prevails i.e. for
nuclear  matter (NM). For that $'$target$'$
quasi-data have  been generated  from finite  $A$
nuclei by taking the large $A$  limit, thus separating surface and volume
contributions $^4$.

The above  does not  touch on  relativity, the  introduction of  which at
first sight  seems to pose  serious difficulties.  Fortunately  there are
mitigating  circumstances in  moderately high  $q$ inclusive  scattering,
which indicate that there is no need for a fully relativistic description
and those seem to offer the possibility of a realistic description of the
above quasi-data. The invoked arguments stress  qualitatively different
roles for  various degrees of  freedom.  Virtually all belong  to passive
spectator nucleons,  which have  a non-relativistic (NR)  average momenta
$\langle p\rangle  \le p_F\ll m$.   Relativity thus enters only  in the
description  of  the  redistribution  by hard  collisions  of  the  large
4-momentum, which  has been acquired  by the  initial  absorption  of a
high-momentum virtual photon.

The proposed  characterization is  actually also  the only  feasible one.
The  very nature  of  a  totally inclusive  (TI)  or semi-inclusive  (SI)
process prescribes the cross section to be an integral over
the $A$-body matrix  element, diagonal in all coordinates  except the one
of the knocked-on nucleon.  However,  with increasing $q$ fewer particles
actually participate in Final State  Interactions (FSI) and frequently it
suffices to retain  density matrices of order $n\le 2$.   Already for the
non-relativistic  case,  a calculation  or  parametrization  is far  from
trivial $^{5,6}$.  It  is difficult to conceive their  determination in a
relativistic theory without making NR connections.

By far the most illustrative and  most thoroughly studied example of such
a  NR  system is  liquid  $^4$He,  for  which  the sketched  program  has
converged  in a  realistic description  $^{7-9}$.  At  first sight  those
studies seem  to be of  little relevance: On all  counts, liquid He  is a
much  simpler system  than  is  NM.  For  current  or foreseen  kinematic
conditions, liquid He is a system of point atoms with an
accurately  known  interaction.  Those  degrees  of  freedom are  neither
internally excited, nor exists there the equivalent of particle
production.

Even noting the  essential differences, it is tempting  to borrow insight
from the relatively simple description of  large $q$ TI scattering to the
case of  NM.  We  recall here  a rather unexpected  result from such an
attempt$^{10}$, stating  that FSI may  be negligible beyond  a relatively
low momentum transfer $q\approx 0.7$ GeV.

In spite of the  above hint that a partially NR treatment  may not be too
far  off the  mark,  the  above model  is  clearly  deficient in  several
respects, mainly in
the use  of a $t$-matrix, non-relativistically generated
by a  potential $V$.  A  relativistic analog could  be a $t$-matrix  in a
Bethe-Salpeter  equation,  driven by  a  mechanism,  which describes  the
dynamics  of the  redistribution  of the  momentum  transfer $q$  amongst
several particles. \footnote * {In this context it is amusing to recall a
truly ancient  piece of  research by L${\rm\acute  e}$vy and  Sucher, who
summed a large class of diagrams in a bona fide field theory.  The result
has  the appearance  of  the  covariant analogue  of  a standard  eikonal
amplitude,    now    generated    by    a    generalized    4-dimensional
potential.$^{11}$} Simultaneously it should be flexible enough to also
account of production  channels.

Having  outlined  the   topic  and  the  expected   difficulties  of  its
description, we review some literature  and start with Butler and Koonin,
who   calculated   particle-hole    scattering   contributions   to   the
total
response$^{12}$.  Part of  those can be indeed be shown  to belong to the
dominant high-$q$ FSI.

Next there are on-going studies of  TI scattering starting from the Plane
Wave Impulse Approximation (PWIA)$^{13}$, where the recoiling particle is
assumed to  be too  fast to  interact with  the core  or medium.   Only a
particular  single-particle property,  namely  the  NR spectral  function
$^{14,15}$ governs the process, while  relativity   enters only through
kinematics  of  the knocked-out particle.  It  is  clearly  necessary  to
assess  the   importance  of FSI
contributions.   One such  attempt is  due to  Benhar $et\,al^{16}$,  who
postulated a form for  FSI, which in the end has  elements in common with
the cited  theoretical framework  for liquid He. After having developed
our expression for the FSI, a comparison is clearly called for. We also
mention an
approach by Frankfurt $et\,al$ who mostly  concentrate on the ratio of TI
cross sections for finite $A$ nuclei and the $D\,^{17}$.

We start in Section  2 with the TI cross section in  terms of the assumed
nuclear current.  We recall that the PWIA provides a natural link between
the associated  $nucleonic$ current tensors  and that cross  section.  It
also  visualizes  the  distinction  between  matrix  elements  where  the
intermediate  is either  a nucleon  or else  any non-strange  $B=1$ state
(Section  3a),  both  of  which  interact with  core  nucleons.   In  the
attempted  description  we   shall  draw  on  rich   experience  with  NR
systems$^{7-9}$.  We  thus review in  Section 3b the asymptotic  limit of
the response as a starting point, the systematic treatment of FSI effects
as decreasing powers of $1/q$ and the first cumulant approximation in the
case of dominant binary $NN$ collisions in FSI.  That section also treats
the incorporation of  the same in the TICS and  an attempted relativistic
generalization.

Whereas the PWIA is a well-defined  and standard starting point, the
theory of the responselacks aconsistent and
systematic treatment of FSI.  Nevertheless, in order  to make a
comparison with  previous work on the  IA, we formulate in  Section 3c a
first  cumulant approximation,  based on  the lowest  order terms  in the
$1/q$  expansion of the IA series. In Section 4  we  only mention
the $'$inelastic$'$  part of  the  TICS: FSI between an
exited knocked-out particle and core nucleons are disregarded.

Selecting only the $'$elastic$'$  (purely nucleonic) contributions of the
TICS, we display in Section 5  scaling plots in the relativistic West and
IA  scaling  parameters  and  discuss  the  manifestly  superior  scaling
properties for the  former.  In Section 6 we report  on numerical results
for the  two versions  for the  first cumulant and  test in  addition the
sensitivity for $NN$ scattering parameters.   In Section 7 we compare the
two sets of predictions with the data and, given the known theoretical
uncertainties, discuss the measure of agreement  one may  expect.
Finally we make  a critical comparison with the IA  approach by Benhar et
al  $^{16}$,   including  remarks   on  the   expected  role   of  colour
transparency.

{\bf 2. The $eA$ inclusive cross section. }

For the cross section for the TI (TICS) reaction
$e+A\rightarrow e'+X$ one has $^{18}$

$${d^2\sigma\over{d\Omega d\epsilon'}}={\alpha^2\over Q^4}
{\epsilon'\over\epsilon}L^{\mu\nu}(k,q)W^A_{\mu\nu}(q),\eqno (2.1)$$
\noindent
where $\alpha$=1/137 and $L^{\mu\nu},W_{\mu\nu}$  are
respectively, the electron and the target current tensors. The first is

$$L^{\mu\nu}(k,q)=2(k^{\mu}k'^{\nu}+k'^{\mu}k^{\nu}-
g^{\mu\nu}(k.k')),\eqno(2.2)$$
\noindent
with $k=(\epsilon,\vec  k)$ and $k'=(\epsilon',\vec k')$, the initial and
final electron 4-momentum; $q=(\omega,\vec q)=k'-k$,
the transferred 4-momentum and $Q^2=\vec q^2-\omega^2=-q^2$.
When expanded in target states $P_{\alpha}$
the nuclear current tensor in the ground state $'P_A'$ reads

$$W^A_{\mu\nu}(P_A,q)=
\sum_{\alpha}\int d^4P_{\alpha}\langle P_A|J^A_{\mu}(0)|P_{\alpha}\rangle
\langle P_{\alpha}|J^A_{\nu}(0)|P_A\rangle \delta^4(P_A-P_{\alpha}+q)
\eqno(2.3)$$

The nuclear current $J^A$ in (2.3) is not uniquely defined. It
depends on the underlying theory which generates the states on which the
current acts. For instance, in standard NR nuclear theories one assumes

$$J^A=\sum_ij_i,\eqno(2.4)$$
\noindent
thus  neglecting many-pa1rticle, non-nucleonic exchange contributions.
In the PWIA one gets $^{19}$
$$W^{A,PWIA}_{\mu\nu}(P_A,q)=(4\pi i)^{-1}{\rm Re}\sum_n\int d^4 p
{|\Gamma_n^r(P_A-p,p)|^2 \over {(P_A-p)^2-
\left (M_{A-1}^{(n)}\right)^2+i\epsilon}}\sum_j w^{(j)}_{\mu\nu}(p,q)
\eqno(2.5)$$
Here $M_{A-1}^{(n)}=(A-1)m+E_n$ is the mass
of the residual core in a state $n$ and

$$w^N_{\mu\nu}(p,q)=\sum_{\beta}\int d^4p_{\beta}
\langle p|j^N_{\mu\nu}(0)|p_{\beta} \rangle
\langle p_{\beta}|j^N_{\mu\nu}(0)|p_A\rangle \delta^4(p-p_{\beta}+q),
\eqno(2.6)$$
\noindent
the off-shell nucleon  current tensor. $\Gamma^r_n$ is  a vertex function
including  the  propagator  of  the struck  particle  and  describes  the
dissociation  of the  target in  a  core in  a  state $n$  and the  above
nucleon. Since $p^2 \ne m^2$, Eq. (2.5) links the current tensor of the
on-shell nucleus ($P_A.P_A=M_A^2)$ to the same for an off-shell
constituent (Fig. 1). This fact clearly
prevents a simple relation between  the two current
tensors.  However, assuming  the $p_0$ integral in (2.5)  to be dominated
by the contribution of the pole at $p_0=m-E_n$, the target current tensor
there for symmetric nuclear matter can be written as
$$\eqalignno{W^{A,PWIA}_{\mu\nu}(P_A,q)&\approx
 A/2\int d\vec p\int dE P_h(p,E)
 [w^p_{\mu\nu}(\vec p,p_0,q)+w^n_{\mu\nu}(\vec p,p_0,q)]&(2.7a)\cr
 P_h(p,E)&\equiv \sum_n \mid \Gamma_n^{nr}(\vec p)\mid^2
 \delta(E+E_n), &(2.7b)\cr}$$
\noindent
where the on mass-shell NR vertex function $\Gamma^{nr}$
(with additional energy factors) replaces the relativistic $\Gamma$
in (2.5). The above result further features
the NR nucleon-hole spectral function $P_h(p,E)$,
which measures the probability to remove from the target in its
ground state a nucleon with momentum $\vec p$, leaving the core with
excitation energy $E$; for nuclear matter it will be
assumed to be identical for $p$ and $n$.  As in (2.5), the target in
$W^A(P_A,q)$ above is on its mass shell
, but this is
not the case with the constituent nucleons in $w^N(p,q)$.

Whereas the  struck particle in (2.7a)  is a nucleon, the  particle after
recoil may be any non-strange $B$=1 one:  we shall separately discuss the
corresponding  $'$elastic$'$  and  $'$inelastic$'$ contributions to the
nucleon current tensor $w_{\mu\nu}\,^{20}$.

{\bf 3. The $'$elastic$'$ (nucleonic) part of $w_{\mu\nu}$.}

{\bf 3a. Plane Wave Impulse Approximation}

Since the nuclear current tensor is most directly related to the nucleon
tensor in the PWIA, it is natural to first discuss that approximation.
Substituting  (2.2) and (2.7a) into  (2.1), one finds
for the 'elastic' contribution to the TICS per nucleon
$\left(e_0(p)=\sqrt{\vec p^2 +m^2}; \hat q=\vec q/|q|\right)$
$$\eqalignno{
\left({d^2\sigma^{el}_{eA}\over{d\Omega d\epsilon'}}\right)&^{PWIA}
=\int \p3\int dE P_h(p,E)
{d^2\bar\sigma^{el}_{e\langle N\rangle}\over{d\Omega d\epsilon'}}
\delta(\omega-E+m-\ee) & (3.1)\cr
{d^2\bar\sigma^{el}_{e\langle N\rangle}\over{d\Omega d\epsilon'}}
&={1\over 2}\sum_{N=p,n}{d^2\bar\sigma^{el}_{eN}\over{d\Omega
d\epsilon'}} & \cr
{d^2\bar\sigma^{el}_{eN}\over{d\Omega d\epsilon'}}
&=\left({d\sigma\over {d\Omega}}\right)_M{m^2\over{e_0(\vec p)\ee}}
\left[T_1(Q,p,\theta)\bar w_1^{N(el)}(p,q)+
T_2'(Q,p,\hat p.\hat q,\theta)\bar w_2^{N(el)}(p,q)\right],& (3.2)\cr}$$
\noindent
where $\tilde \sigma_{eN}^{el}$ and  $\sigma_M$ are the off-shell $eN$
TI and  Mott cross sections.
\footnote* {A bar over a quantity (cf. Eq. (3.1)
indicates that  a $\delta$-function has been removed.}
When both spectator core and recoiling nucleon  are taken on
shell,  energy   conservation  imposed  by  the   above  pole  dominance
determines the energy of the struck nucleon in two ways

$$p_0=\ee-\omega=m-E\ne e_0(\vec p).\eqno(3.3)$$
\noindent
There is no unique prescription to exactly incorporate  the  measure
$\tilde \omega-\omega\equiv p_0-e_0(p)$ to which the above nucleon is
off-shell. In one procedure $^{21,22}$ one replaces
$Q^2$ by $\tilde Q^2=\vec q^2-{\tilde \omega}^2$, where
$$\tilde\omega=\ee-e_0(p)\eqno(3.4)$$
\noindent
This change is selectively applied
in the kinematic and structure factors in (3.2).
Those then read ($\vec p_{\perp}=\vec p-\vec p.\hat q$)

$$\eqalign{
T_1&=2{\tilde Q^2\over{Q^2}}{\rm tan}^2{\theta\over 2}+
{Q^2\over{|\vec q|^2}}\left({\tilde Q^2\over {Q^2}}-1\right)\cr
T_2'&=\left({\vec p_{\perp}\over m}\right)^2\left({\rm tan}^2
{\theta\over 2}+{Q^2\over{2\vec q^2}}\right)+
{Q^4\over{\vec q^4}}\left({e_0(\vec p)+\ee\over{2m}}\right)^2}
\eqno(3.5)$$
\noindent
The (reduced) elastic nucleon structure functions have their standard
forms in terms of the electric and magnetic form factors
$\left(\eta=Q^2/4m^2,\tilde \eta=\tilde Q^2/4m^2\right)$

$$\eqalign{
\bar w_1^{N(el)}&=\eta [G^N_M(Q^2)]^2\cr
\bar w_2^{N(el)}&={1\over {1+\eta}}
[(G^N_E(Q^2))^2+\eta(G^N_M(Q^2))^2]+{\tilde\eta-\eta\over {(1+\eta)^2}}
[(G^N_E(Q^2))^2-(G^N_M(Q^2))^2]}\eqno(3.6)$$

Eq. (3.1)  shows that the $eA$ TICS in PWIA is the same for $eN$,
weighted by the nucleon-hole psectral function. Extraction of the $eN$
TICS at some  given values for  $p,E$ leaves the current charge response
in  PWIA.  Virtually all efforts to determine FSI corrections have
until now focussed on responses and we shall investigate below an
algorithm which leads to the same for the TICS.

{\bf 3b. FSI in first cumulant approximation: Generalized GRS theory.}

There  does not  exist a  reliable  relativistic theory  for a  many-body
system and consequently
the same holds for its response. Fortunately not all
features of the high $q,\omega$ response are relativistic. For instance,
momenta of  nucleons in  the target  will rarely  exceed the  NR Fermi
momentum. In contradistinction the encounters of the
knocked-out proton  and the above  target nucleons
are collisions at  high lab momentum,  where
particles could be created or absorbed and which processes cannot
be described in a NR theory.
It thus seems natural to treat separately  hard processes ($'$hard$'$ on
the hadronic scale) and NR 'low'-energy processes. In the development
we shall be guided by the NR theory of Gersch,
Rodriguez and Smith (GRS)$^5$, which we now review. Its salient feature
is a series expansion in $1/q$ of the (reduced) response \footnote*
{The  actual expansion parameter in (3.7) is  not $m/q$, but $q_0/q$
with $q_0$  some momentum,typical of the interaction $V\,^8$.}
$$\phi(q,y)\equiv(q/m)S(q,\omega)=\sum_{l\ge 0}(m/q)^lF_l(y),\eqno(3.7)$$
where the energy transfer $\omega$ is replaced by an equivalent kinematic
variable $^{5,23}$
$$y\rightarrow y^{nr}=\(y_{GW}^{nr}=\){m\over {|\vec q|}}
\left(\omega-{|\vec q|^2\over{2m}}\right)\eqno(3.8)$$
The above reduced response has as asymptotic limit
and dominant FSI correction, which for infinitely extended matter read
$$\eqalignno{
F_0(y)&=\int \p3 n(p)\delta(y-p_z)=
(4\pi^2)^{-1}\int_{|y|}^{\infty}dp p n(p)& (3.9a)\cr
(m/q)F_1(y)&=(2\pi)^{-1}{\rm Re}
\int_{-\infty}^{\infty}ds e^{iys}\int d\vec r
\rho_2(\vec r-s\hat q,0;\vec r,0)\(i\tilde\chi_q^{1,2}(\vec r,s)\)& \cr
\tilde\chi_q^{1,2}(s,\vec r)&=(m/q)\int_0^s d\sigma
\(V(\vec r-\sigma\hat q)-V(\vec r -s\hat q)\), &(3.9b)\cr}$$
\noindent
where $'1,2'$ refers to the two parts in (3.9b).
$V$  is  the interaction  between  the  constituents, $n(p)$  the  single
particle momentum  distribution and $\rho_2(1,2;1'2)$,  the half-diagonal
2-particle  density matrix. Defining  a  non-diagonal pair  distribution
function through ($\vec r=\vec r_1-\vec r_2)$)
$$\rho_2(\vec r_1,\vec r_2;\vec r_1',\vec r_2)\rightarrow
\rho_2(\vec r+s\hat q,0;\vec r,0)=\rho\rho_1(s,0)\zeta_2(\vec r,s)
\eqno(3.10)$$
In terms of the standard pair-distribution function $g(r)$, the
approximation which we shall use  reads $^5$
$$\zeta_2(\vec r,s)\approx \sqrt{g(\vec r+s\hat q)g(r)}\eqno(3.11)$$

Using the fact that an
inclusive process is by assumption
induced by a single-particle operator
Eq. (2.4), one can prove that the response in the GRS theory can
generally be written as a convolution of the asymptotic response
and a FSI factor$^5$. Thus with $e_0^{nr}(p)= p^2/2m$
$$\eqalignno{
S(q,\omega)^{as,nr}&=\int \p3 n(p)\delta\(\omega+\ff-\ee\) & (3.12a)\cr
S(q,\omega)^{FIS,nr}&=\int d\omega'(m/q)R_q^{nr}\(m/q)(\omega-\omega')\)
S(q,\omega')^{as,nr} & \cr
&=\int \p3 n(p)(m/q)R_q\((m/q)(\omega+\ff-\ee)\) & (3.12b)\cr}$$
Or replacing $\omega\rightarrow y$, Eq. (3.8)
$$\phi(q,y)=\int dy'R_q(y-y')\phi(q,y')^{as}=\int dy'R_q(y-y')F_0(y')
\eqno(3.13)$$

It becomes increasingly difficult
to calculate the coefficient functions $F_l(y)$ in (3.7) beyond the
lowest orders (3.9a), (3.9b), and the same  holds for the general FSI
factor $R$ in (3.12). However, for sufficiently high momentum and energy
losses $(q,\omega)$, not too far from the QEP at $\omega=q^2/2m$ one may
use an approximation, retaining only binary collisions between the
recoiling and core nucleons. Their  effect is  contained in the  first
cumulant approximation of  $R_q(\omega)$ $^{24,25}$. For
infinitely extended matter it reads
$$\eqalignno{\tilde R_q(s)&=\int_{-\infty}^{\infty} dk/(2\pi)
e^{iks} R_q(k)=e^{\tilde\Omega_q^{1,2}(s)} & (3.14)\cr
\tilde\Omega_q^{1,2}(s)&=\(\rho\int d\vec r\zeta_2(\vec r,s)
\tilde\Gamma_q^{(1,2)}(\vec r,s)\)
& (3.15a)\cr
\tilde\Gamma^{(1,2)}&={\rm exp}\(\tilde\chi^{(1,2)}\)-1
=\tilde \Gamma^{(1)}+\tilde\Gamma^{(2)}+\tilde\Gamma^{(1)}
\tilde\Gamma^{(2)},& (3.15b)\cr}$$
\noindent
where the superscripts on the FSI factors refer to the two terms in the
integrand of (3.9b). For moderate $\Omega$ one may expand
the exponent in (3.15a) and the lowest order terms
$\tilde R\approx 1+\tilde\Omega$ coincide with (3.9).

The discussion till now focussed on the  response (3.12) and not on the
required TICS. However, one notices that  the expression for the
asymptotic limit of the
response contains  a $\delta$-function with  argument, identical to the
one in the TICS, Eq. (3.1) (cf. (3.9a) and (3.12a)). It
is then plausible that TICS including FSI effects are generated by a
similar replacement of the above $\delta$-function by $R$,  thus
$$\eqalign{
\left ({d^2\sigma^{el}_{eA}\over {d\Omega d\epsilon'}}\right)^{nr}
&\rightarrow
\int \p3 n(p){d^2\bar\sigma^{el}_{e\langle N\rangle}\over{d\Omega
d\epsilon'}} (m/q)R_q^{nr}\(m/q)\(\omega+\ff-\ee\)\)\cr
&=\int \p3 n(p){d^2\bar\sigma^{el}_{e\langle N\rangle}\over{d\Omega
d\epsilon'}} (m/q)R_q^{nr}(y^{nr}-p_z),}\eqno(3.16)$$

After the above outline of a NR approach, we endeavour a
generalization to $q,\omega$ processes in the few-GeV range.
Relativistic kinematics first influences the form of
the scaling variable. The NR GRS-West variable (3.8) is thus
generalized to $^{26}$
$$y\rightarrow y^r_W={m\over {|\vec q|}}\left(\omega-{Q^2\over {2m}}
\right),\eqno(3.8')$$
The relativistic analogue of the asymptotic limit and the TICS including
FSI then become (cf. Eqs. (3.12a), (3.12b) and (3.16))
$$\eqalignno{
\left ({d^2\sigma^{el}_{eA}\over{d\Omega d\epsilon'}}\right )^{as,r\,W}
&=\int\p3 n(p){d^2\bar\sigma^{el}_{e\langle N\rangle}\over{d\Omega
d\epsilon'}}
\delta\left(\omega-{Q^2\over {2m}}-{\vec p\vec q\over m}\right)
& (3.17a)\cr
\left ({d^2\sigma^{el}_{eA}\over{d\Omega d\epsilon'}}\right )^{FSI,r\,W}
&=\int \p3  n(p){d^2\bar\sigma^{el}_{e\langle N\rangle}
\over{d\Omega d\epsilon'}}
(m/q)R^{r\,W}_q\left \((m/q)\left(\omega-{Q^2\over{2m}}-
{\vec p\vec q\over{m}}\right)\right\)&\cr
&=\int \p3  n(p){d^2\bar\sigma^{el}_{e\langle N\rangle}
\over{d\Omega d\epsilon'}}(m/q)R^{r\,W}_q(y^r_W-p_z) &(3.17b)\cr}$$
\par
We argued above that, when using the first cumulant approximation (3.14),
one may continue to use  NR quantities as are $n(p),\rho_2$. This is not
the case with the off-shell profile $\tilde\Gamma$, Eq. (3.9b) in terms
of a static, energy-independent $V$. Such a single channel description
(ora  many-channel analog) is unsuited to describe
particle production and absorption in hard collisions required
in Eq. (3.15b).

In a standard way one avoids this problem by replacing $V$ by an
effective interaction which is of short range or,
equivalently, has a Fourier transform which depends only on the
longitudinal momentum transfer $\vec Q_{\perp}$. Thus for the off-shell
profile due to the first interaction in (3.9b)
$$\eqalignno{\tilde\Gamma^1_q(\vec r,s)&\equiv
e^{i\tilde\chi^{(1)}(\vec r,s)}-1
\approx ie(q)/q\int d^3Q/(2\pi)^{-3}\int_0^s d\sigma\, {\rm exp}
\(iQ_z(\zeta-z)\,{\rm exp}\(i\vec Q_{\perp}\vec b\)v_{eff}(Q_{\perp})
& (3.18a)\cr
&\approx\theta(s-z)\theta(z)\Gamma_q^{(1)}(b)& (3.18b)\cr
\Gamma_q^{(1)}(b)&=e^{i\chi_q^{(1)}(b)}-1=
{2\pi i\over q}\int d^2Q_{\perp}/(2\pi)^2
e^{i\vec Q_{\perp}\vec b}f_q(\vec Q_{\perp})& (3.18c)\cr}$$
\noindent
with $\Gamma$ the $on-shell$ profile.
The choice $v_{eff}(\vec Q_{\perp})\to t(\vec Q_{\perp})$, produces the
relation Eq. (3.18b) between the off-shell profile and phase and
their on-shell analogs. The latter in turn are expressed in terms of
(spin-isospin averaged) elastic $pN$ amplitude which, if of
diffractive nature, can approximately be expressed as
($\tau_q$=Re$f_q(0^{\circ})/$Im$f_q(0^{\circ})$)
$$\eqalign{f_q(Q_{\perp})&=f_q(0^{\circ})\tilde a_q(Q_{\perp})
\approx f_q(0^{\circ}){\rm exp}[-(Q_{\perp}/Q_q^{(0)}]^2\cr
&\approx\,i{q\over{4\pi}}\(1-i\tau_q\)\sigma_q^{tot}(q)A_q(b)\cr
A_q(b)&=\int d^2\vec Q_{\perp}/(2\pi)^2 {\rm exp}
\(i\vec Q_{\perp}\vec b\)\tilde a_q(\vec Q_{\perp})
\approx ((Q_q^0)^2/4\pi){\rm exp}\(-(bQ_q^0/2)^2\)
\to \delta^{(2)}(\vec b),}\eqno(3.19)$$
\noindent
where the last expression above gives the zero-range limit. Substitution
of (3.19) into (3.18c) gives
$$\tilde\Gamma_q^{(1)}(\vec r,s)=\theta(z)\theta(s-z)(\sigma_q^{tot}/2)
(1-i\tau_q)A_q(b)\eqno(3.20)$$

We now return to the second phase in (3.9b)
which in the non-relativistic case satisfies
$$\tilde\chi^{(2)}(\vec r,s)=
-(\partial /\partial s)\tilde\chi^1(\vec r,s),\eqno(3.21)$$
and which form  we assume to hold also for hard collisions, thus
$$\tilde\Gamma^{(2})(\vec r,s)\approx
-s(\partial/\partial s)\tilde\Gamma^{(2)}(\vec r,s)=
s\theta(z)\delta(s-z)(\sigma^{tot}/2)(1-i\tau)A(b)\eqno(3.22)$$
The product term in (3.15b) can be written as
$$\eqalign{
\tilde\Gamma^{(1)}\tilde\Gamma^{(2)}&\approx-(1/2)
(\partial /\partial s)\theta(s)
(1-i\tau)^2 \(A(b)\)^2\cr
&\approx-(1/2)s\theta(z)\delta(s-z){(1-i\tau)^2 \o {1+\tau^2}}
\sigma^{tot,el}\(A(b)\)^2,}\eqno(3.23)$$
where use has been made of
$$\left\(\sigma^{tot}/2)A(b)\right\)^2={\sigma^{part,el}\o {1+\tau^2}}=
{A(b)\o {1+\tau^2}}\sigma^{tot,el}\eqno(3.24)$$
The required partial elastic cross section in impact parameter
space $\sigma^{part,el}(b)\equiv A(b)\sigma^{tot,el}$ manifestly
satisfies $\sigma^{tot,el}=\int d^2\vec b\,\sigma^{part,el}(b)$.
Using (3.20), (3.22) and (3.23) one finds an approximate parametrization
for the total off-shell profile (3.15b) and the FSI factor for NM
$$\eqalign{
\tilde R_q^{r,W}q(s)&={\rm exp}\(\tilde\Omega_q^{r,W}q(s)\)\cr
\tilde\Omega_q^{r,W}(\vec r,s)&=
-\rho {\sigma_q^{tot}\o 2}
(1-i\tau_q)\int d^2\vec b A_q(b)\left\lbrace\int_0^s dz\zeta_2(bz,s)dz
-{s\zeta_2(bs,s)\o {1+i\tau_q}}\(i\tau_q+\sigma_q^{reac}/\sigma_q^{tot}
\right\rbrace }\eqno(3.25)$$
Eq. (3.25), the first cumulant approximation to the TICS in
the generalized GRS theory, is our major result. Substitution of its
Fourier transform (3.14) in (3.17) provides the $'$elastic$'$ FSI part
to the TICS.

One notices that besides the total $pN$ cross section, Eq. (3.25)
features the total reactive cross section, which also emerged in
recent studies of high energy nucleon $(e,e'p),(p,2p)$
knock-out reactions $^{27,28}$. Their presence
accounts for $NN$ induced processes with
particle production and absorption in intermediate states.

Next we remark  on the expected $q$  dependence.  We recall that  in a NR
theory  the  dominant  FSI  (3.9)  in terms  of  an  interaction  $V$  is
$\Oscr(q^{-1})$.  This  is no more  the case after the  replacement $V\to
V_{eff}\propto f$ and subsequent parametrizations (3.20), (3.22) and
(3.23) of $f\propto q\sigma^{tot}$, which causes cancelation of the
explicit $1/q$ factor. The  remaining $q$-dependence, implicit in
cross sections,  slope parameters, etc., is a moderate one.
The same  is then the  case for $\tilde\Gamma_q(s),\tilde\Omega_q(s).$
\footnote*
{Note  that  the  profile increases  linearly  with $s$,  which
ensures convergence of the first  cumulant approximation (3.15a), but not
of its expansion, of which (3.9)  gives the lowest order terms.  This can
be  understood once  it is  realized that  $s$ is  the path  distance the
recoil particle  traverses and which for  NM is unlimited.  All  terms in
the GRS  series (3.8) are  weighted by the  non-diagonal, single-particle
density $\rho_1(s,0)$ (cf.  Eqs. (3.9a), (3.10)), which however is unable
to suppress powers of $s$ to all orders.
Actual dimensions of finite nuclei limit the distance the knocked-out
proton can  recoil.}

A  last   remark  here   regards  the  relation   between  semi-inclusive
$(e,e'p)\,^{27,28}$  and  total-inclusive  $(e,e')$  reactions.   To  the
extend that the former  is dominantly single-nucleon induced, integration
over the outgoing  proton energy and scattering angle  should produce the
TI cross section.  It is indeed of interest to investigate this 'sumrule'
for finite nuclei: the total yield of knocked-out protons in NM vanishes,
since those have to pass an infinite path-length.

{\bf 3c. FSI in first cumulant approximation: IA version.}

In Section 3b above we developed a theory of inclusive scattering
for $Q^2$ of several GeV, trying to incorporate minimal
relativistic adjustments in the NR GRS theory for FSI. The latter
is non-perturbative in the interaction $V$ and permits the isolation
of the asymptotic limit of the response as well as of dominant
corrections to it. These depend on  a scaling variable $y_{GW}$, which
the theory naturally selects$^5$.

The GRS theory is not the only candidate for a NR
description of the nuclear response. Frequently one uses the IA
series which is a perturbation theory in the interaction between the
knocked-out nucleon and the core. The lowest order contribution
produces the well-known expression in terms of the single-nucleon
spectral function$^{13}$ (cf. for instance Eq. (2.7))
and the corresponding reduced response can be given in terms of
the energy loss or of a corresponding IA scaling variable
$$y_0^{nr}=-q+\sqrt{2m(\omega-\langle\Delta\rangle)}\eqno(3.26)$$
Aside kinematic variables, $y_0$ features an average separation
energy $\langle \Delta\rangle$, which naturally constitutes a
complication.

To our knowledge there exist no evaluation of FSI
in powers of the above-mentioned, many-particle residual interaction, as
a systematic IA expansion of the response demands. The latter is
not in any way parallel to the same in the GRS theory, and in particular
does not lead to a power series in $1/q$ \footnote *{Nevertheless
the literature contains statements that, starting from the PWIA, FSI
are represented by the GRS series $^{29}$.}
However, it appears possible to isolate in the IS series
terms of the lowest two orders in $1/q$. It has thus been shown that
$$\eqalign{F_0(y_0)&=F_0(y_{GW})\cr
(m/q)F_1(y_0;\(\chi^1\))&=(m/q)F_1(y_{GW};\(\chi^{1,2}\)),}\eqno(3.27)$$
\par
\noindent
i.e. the dominant FSI of order $1/q$ is the same as function of $y_0$
as of $y_{GW}$, provided one disregards the second term in
the integrand in (3.9b)$^{30}$. We do not know of a generalization of
(3.27) for $F_l(y_0)\,,l\le 2$. In particular there is no proof that
in IA binary collisions to either the reduced response or the TICS (3.1),
can be represented in the form (3.12b), respectively (3.16).
Nevertheless, in order to enable a comparison with other work we shall
follow this line based on (3.27) and just assume that
the first cumulant is such an approximation to the IA series. Thus
$$\tilde R^{r,IA}(s)\equiv{\rm exp}\(\tilde\Omega^{(1)}(s)\)={\rm exp}
\(\rho\int d\vec r\zeta_2(\vec r,s)\tilde\Gamma^{(1)}(\vec r,s)\)
\eqno (3.28)$$
\noindent
Again, replacing the $\delta$ function in (3.1) by $R$, the Fourier
transform of $\tilde R$ (cf. Eq. (3.14)), one formally
generates the IA version of the first cumulant  for the TICS
$$\eqalignno{
\left ({d^2\sigma^{el}_{eA}\over{d\Omega d\epsilon'}}\right )^{as,r\,IA}
&=\int\p3 n(p){d^2\bar\sigma^{el}_{e\langle N\rangle}\over{d\Omega
d\epsilon'}} \delta
\left(\omega-|\langle\Delta\rangle|+m-\ee)\right\)
& (3.29a)\cr
\left ({d^2\sigma^{el}_{eA}\over{d\Omega d\epsilon'}}\right )^{FSI,r\,IA}
&=\int \p3  n(p){d^2\bar\sigma^{el}_{e\langle N\rangle}
\over{d\Omega d\epsilon'}}
(m/q)R^{r\,IA}_q\left\(
(m/q)\left\(\omega-|\langle\Delta\rangle|+m-\ee\right)\right\)&\cr
&=\int \p3  n(p){d^2\bar\sigma^{el}_{e\langle N\rangle}
\over{d\Omega d\epsilon'}}(m/q)R^{r\,IA}_q(y^r_0-p_z), &(3.29b)\cr}$$
\noindent
The last equation features the relativistic PWIA scaling variable
$$y_0^r=-q+\sqrt{(\omega-\langle\Delta\rangle)^2+2m(\omega-
\langle\Delta\rangle)},\eqno(3.26')$$
which permits to convert (3.29b) to a cross section in $\omega$.

This concludes  the treatment  of the  response inasmuch  as due  to
$'$elastic$'$ processes. We still have to mention $NN$ processes where
the knocked-out proton is not a nucleon.

{\bf 4. $'$Inelastic$'$ parts of $W_{\mu\nu}$}

A treatment of  FSI for the $'$inelastic$'$ part of the nuclear current
tensor and the TICS
requires a description of the  interaction of the non-nucleonic
recoiling particle with a target  $N$. Apparently no effort has been
made, parallel to the developments in Sections 3 for the $'$elastic$'$
part. The former is commonly treated in the  PWIA (3.1a)$^{14,20}$ and
summarizing we  have
$$\left\({d^2\sigma_{eA}\over{d\Omega d\epsilon'}}\right\)^{tot}\approx
\left\({d^2\sigma_{eA}^{el}\over{d\Omega d\epsilon'}}\right\)^{FSI}+
\left\({d^2\sigma^{inel}_{eA}\over{d\Omega d\epsilon'}}\right\)^{PWIA}
\eqno(4.1)$$
This completes the outline of our  procedure: Eqs. (3.17b), (3.25),
(3.8$'$) and (4.1) enable  a calculation  of the TICS including  FSI to
the described level in the GRS version. The same for the IA uses Eqs.
(3.29b), (3.28), (3.26$'$), and (4.1).

{\bf 5. $y$-scaling}

Before reporting calculations we insert here a
discussion of an issue, entirely based on data and some theoretical
concepts, and emphatically not on numerical results. Assuming here
that  one   can  factor  out an  average $e\langle N\rangle$ cross
section  from the  TICS  (3.17b) or  (3.29b) the  remaining
factor  dependent on  incident  electron  energy $\epsilon$,  scattering
angle $\theta$ and  energy loss $\omega$, can be converted  to a function
of the 3-momentum and  a relativistic scaling variable. Doing so
at  $p_z=y^r$, the smallest possible transverse momentum of the
knocked-on nucleon compatible with energy conservation,
one finds for the so-called scaling function
$$\eqalignno{G(q,y)&
\equiv K^{-1}d^2\sigma_{eA}/d^2\sigma_{e \langle N\rangle} & (5.1)\cr
K_W&=|(\partial \omega/\partial p_z|^{-1}_{p_z=y_W^r}=
\sqrt{e_0^2(q)+2qy_W^r}/q & (5.2a)\cr
K_0&=|(\partial \omega/\partial p_z|^{-1}_{p_z=y_0^r}=
\sqrt{e_0^2(q)+2qy_0^r+(y_0^r)^2}/(q+y_0^r) &(5.2b)\cr}$$
The left hand  side  of  Eq. (5.1) represents direct  experimental
information,  related to  the  reduced  response $G(q,y^r)$ and
general knowledge on its behaviour in $q$ is limited to the
$'$elastic$'$ parts. One thus routinely removes
from data points $'$inelastic$'$ parts discussed in  Section 5,
and concentrates on $G^{'el'}(q,y)$.

One frequently observes bunching of $G^{el}$ data over
some  $y$ range. If the spread
$$\Delta G^{el}(q,y)=G^{el}(q,y)-G^{el,\infty}(y)$$
decreases with $q$, one has  for all practical purposes
reached the asymptotic or  scaling limit $G^{el,\infty}$: Beyond
$q\ge q^{as}$  FSI are by
definition small. We remark that the same is sometimes said for bunching
within a $G$-width of the order, or larger the the above scaling limit.
Since the  $q$-dependent FSI is then of the same order, or
larger than the
asymptotic response,  it seems more appropriate  to use in that  case the
term coarse scaling.

In  the above  no mention  has been  made of a particular or preferred
scaling variable  $y$. Deviations  from ideal  scaling
clearly depend on the choice of  $y$  and
one  faces the  question which scaling  variable is
best, i.e. which one minimizes FSI over a  certain $y$  interval
$^{35}$.  Again, the data without interference of any theory, decide the
outcome.

We recall that  since in (3.9b) $\chi^{(2)}\ll\chi^{(1)}$, the FSI factor
$\tilde R$  in the  first cumulant  approximation (3.25)  in
either the  GRS or IA  approach, discussed in Sections 3,4 are quite
similar. As a consequence, the theory of Sections 3a and 3b predict
essentially the same TICS as function  of either scaling  variable
$y_W$ or $y_0$, but not as function of $\epsilon,\theta$ and $\omega$.
It is clearly of interest to  compare the scaling plots  for those
scaling variables.

A glance on Figs. 4,5 (cf. Fig. 11 in Ref. 14 for a similar scaling
plot in $y_0$) shows that the spread in the scaling plot as
function of $y_W^{rel}$ is considerable smaller than for $y_0^{rel}$.
Differently stated, the data show that FSI in the West variable are
smaller than the same in the IA variable. This result is by itself an
incentive to prefer the GRS theory which naturally accommodates $y_W$

\par
{\bf 6. Numerical results}
\par

We start mentioning the following input elements:

i) The single nucleon momentum distribution for realistic $NN$ forces
$^{32}$.

ii) Dipole $N$ form factors are  as in Ref. 33.

iii) Parameters, collected in Table I and
characterizing the representation
(3.19) of diffractive, elastic $pN$ amplitudes. Appropriate to symmetric
NM, average $pp,pn$ amplitudes have been used throughout.

iv) $'$Inelastic$'$ cross sections, which have been recalculated
in PWIA by the authors of Ref. 1.

With the above input we then computed from Eqs. (3.25), (3.14) and (4.1)
the TICS of electrons from nuclear matter as function of momentum and
energy transfer $q,\omega$. Computations have
been made for a sufficiently fine $q$ grid in order to  cover
the data points of Ref. (1) as function of incident
electron energy $\epsilon_{el}$,  scattering angle $\theta$ and
energy loss $\omega$. The kinematic conditions are:
$\epsilon_{el}$=2.0 GeV, $\theta= 15^{\circ}, 20^{\circ}$;
$\epsilon_{el}$=3.6 GeV, $\theta= 15^{\circ}, 20^{\circ}, 25^{\circ},
30^{\circ}$; $\epsilon_{el}$=4.0 GeV, $\theta= 30^{\circ}$,
all for variable $\omega$ ranges. The
3- and 4- momenta corresponding to the above kinematics are
4.8$\ge\,q^2$(GeV$^2)\ge $0.025, respectively
3.35$\ge\,Q^2$(GeV$^2)\ge $0.02.

For    $\epsilon_{el}$=2.02    GeV     and    $\epsilon_{el}$=3.6    GeV,
$\theta=15^{\circ}$ the calculated non-nucleonic contributions are by the
authors judged to be much less reliable than for higher energies $^{35}$.
We thus disregard  higher $\omega$ parts where  the $'$inelastic$'$ cross
sections are important or dominate.

The quasi-data$^4$
and our predictions are assembled in Figs. 2a-2g.  The lower
dashed curves give the asymptotic limit of the $'$elastic$'$ part (3.17a)
and show  the characteristic  NM discontinuities corresponding  to $p=\pm
p_F$.  The  upper dashed curves have  $'$inelastic$'$ contributions added
to  it, while  fully drawn  curves  represent results  that include  FSI
effects in $'$elastic$'$ parts (Eq. (3.17b)).

FSI virtually always increase the asymptotic TICS,  decrease as expected
with increasing energy and, under otherwise identical kinematic conditions
grow  with decreasing $\omega$. Generally speaking FSI are relatively
small: Relatively large  effects occur locally in the neighbourhood of
the Fermi surface, where they reflect the characteristic discontinuities
in the asymptotic results.

At this point we mention that the parametrization (3.21)-(3.24) leads to
$\chi^{(2)}\ll\chi^{(1)}$ in (3.9b) and that  consequently, the second
term in the curly brackets in (3.25) is negligible. This is easily
understood, since the range $Q_0^{-1}$
of the elastic amplitude, which is actually  decreasing with $q$,
is always smaller than the effective path length $s_{eff}$
of the knocked-out nucleon. For
vanishing range the non-diagonal pair-distribution $\zeta_2(bs,s)
\to\zeta_2(0s,s)\propto\sqrt{g(0)}$ vanishes in the region of a
short-range strong repulsion, causing $\chi^(2)\rightarrow 0$ .

Below the  Quasi-Elastic Peak  (QEP) in the  elastic part,  the predicted
TICS  is always  larger  than the  observed one, and  follows for  fixed
$\epsilon$ the experimentally observed  strong variations with $\theta$.
At  the high  $\omega$ side  of  the QEP  the  TICS is  dominated by  the
inelastic component of the TICS, Eq. (4.1).

Figs. 3a-g represent results of  TICS calculations from (3.29b) which are
based on (3.28), the IA version  of the first cumulant approximation.  We
have tested average  separation energy $\langle \Delta\rangle$  0, 30 and
60  MeV  which  gives  cross sections,  approximately  shifted  by  those
amounts.  For comparison we also entered  in Figs. 3 full predictions for
the GRS  version, Figs. 2.  We recall that
only for $\langle\Delta\rangle$=0
do the  QEP coincide  and is  a direct comparison  feasible. In  the low
energy  loss  region where  the  TICS  has  its sharpest  variation  with
$\omega$, a non-vanishing average  separation energy
$\langle\Delta\rangle$ causes  shifts with considerable  relative changes
in the cross section (cf.  Figs. 3a-c).   We shall return in Section 7 to
attempts to include FSI in the standard IA.

One naturally wishes  to study sensitivity of the  predictions for input,
primarily for the on-shell $NN$ amplitude.  Although some of the involved
parameters are not  accurately known, there is in  general little leeway.
Yet we thought it  of interest to make a few tests in  order to see their
influence on the low-$\omega$ part of  the TICS. We regard those tests as
a measure  of expected changes,  notably due to the treatment of
off-shell $NN$ scattering: Those  we are unable to really calculate at
present.

We thus  checked for both first  cumulant versions TICS the  influence of
neglecting the  real part  of the  $NN$ amplitude,  its range  $1/Q_0$ or
both.  Putting the ratio $\tau\to 0$ hardly influences the results, but a
vanishing range affects the TICS in  the two versions for all $\omega$ in
nearly identical fashion. In particular for the smallest  $\omega$.
changes as large as a factor $\approx 2$ may occur.  Curves
in Figs. 2a, 2g, 3a, 3g marked by $++$,
correspond to results of the above test.

\par  {\bf 7.   Comparison with  data and  previous approaches.   General
discussion} \par

We have presented above a reasonably realistic description of TI electron
scattering in  the few-GeV region.   Data exist  on a variety  of nuclear
targets   and   range    over   4$\ge   \epsilon_{el}$(GeV)$\ge   $2;
30$^{\circ}\ge\theta\ge  16^{\circ}$ and  energy  losses $\omega\le$  2
GeV.  Since considerable simplification results if translational symmetry
prevails, we first addressed NM data.   By extrapolation to the large $A$
limit  Day $et\,al\,^4$  have generated  from the  above, quasi-data  for
symmetric nuclear matter $^1$ and those we tried to describe.

The starting point  of our approach is  the link in PWIA  between, on the
one hand  the nuclear and the  nucleon current tensors, and  on the other
hand, between  the physical $eA$  and the off-shell $eN$  inclusive cross
sections. We mentioned there
processes, where  the knocked-out particle  is either a
nucleon or  a different non-strange  $B$=1 state ($N$  resonance; $N,\pi$
etc). For the  second  category we  utilized
calculations in PWIA without additional FSI.

In order to describe $'$elastic$"$ cross sections
 beyond  the above  lowest  order, we  turned  to known  non-relativistic
decriptions.  We thus  first reviewed the GRS theory for  the NR response
and recalled, that there FSI are expediently calculated starting from the
asymptotic limit.  Those  FSI can be shown to be  contained in FSI factor
$R$, which appears  convoluted with
an energy-conserving $\delta$-function in the asymptotic response (3.16)
$^{24}$.   The  fact that  a  $\delta$-function  with the  same  argument
appears  in  the  'asymptotic'  limit then  suggests  an  algorithm  for
incorporating FSI also in the TICS.

Since one  is dealing with  relatively high $q$-data, we  represented the
FSI factor $R$ by  its first cumulant  approximation  $^{24}$,
which  describes  binary  collisions  between the  recoiling  and  medium
nucleons. We then suggested a  relativistic generalization of the GRS
theory keeping some NR features. Those follow when noting the different
roles played by  particles in the sysytem.  Thus for
nucleons,  not   active  in  the   propagation  and  scattering   of  the
high-energy, knocked-out  nucleon,  the average  momentum of  a core
nucleon $\langle  p\rangle\le p_F\ll  m$. As a result
one  may continue  to use
non-relativistically calculated basic quantities. In the first cumulant
approximation those  are the  momentum distribution and  the non-diagonal
2-particle density matrix of the system in its ground state.

Next one  has to account  for, generally off-shell $NN$  binary collision
between a knocked-out and a medium nucleon and which takes place at a lab
momentum of several  GeV.  The description of this  hard collision should
be relativistic, but no simple theory can provide the required scattering
amplitude.

Predictions have been confronted with the above mentioned quasi-data.
For  given $\epsilon_{el},\theta$,  one distinguishes  two, qualitatively
distinct    energy   loss regions.    Below   the    quasi-elastic   peak
$\omega_{QEP}=Q^2/2m$,  the TICS  is  largely due  to $'$elastic$'$  $NN$
collisions.   Beyond the  QEP $'$inelastic$'$ processes  grow in
importance and eventually dominate the TICS.  Our
major interest concerns the lower $\omega\,\,'$elastic$'$ parts.

For  the  GRS  version  the   agreement  with  the  data  is  reasonable.
Disagreements occur  mainly for  small $\omega$,  where FSI  increase the
asymptotic TICS, which itself already overestimates the data.  In general
FSI are relatively small, as is  also evident from the scaling plot, Fig.
4.  In addition there is  some disagreement for $\omega\gg \omega_{QEP}$,
where the  computed $'$elastic$'$  contribution is insufficient  to raise
the total TICS to the data.

We  stressed above  that there  is no  reason to  assume the  approximate
validity of an IA version of the
first cumulant, based on  the two lowest order IA
terms in  $1/q$.  Nevertheless we  were curious to  see the results  of a
comparison between the  two versions.  Except for  $\epsilon$=2.0 GeV the
predictions  for low  $\omega$ also overestimate  the  data and  in
addition underestimate the large $\omega$  $'$elastic$'$ part of the TICS
(see comparison below).

In general the first IA cumulant  appears to fit the data somewhat better
than the well-founded  GRS version and does so for  an acceptable average
separation energy of the order of  30-60 MeV.  The above outcome does not
contradict the experimental results for the $experimental$ scaling plots,
discussed in Section 6: It points  at deficiencies in the application and
approximation of the GRS $theory$.In particular we suspect that theNR
relation (3.21) for the two composing phases cannot be generalized to the
relativistic regime.

At this  point we  bring up the  actual IA.  The  lowest order term
has extensively been used in the past, but little progress has beenmade
on a systematic treatment of FSI beyond the PWIA. An exception is work
by Benhar  $et\,  al\,$  $^{14}$. However, their computation of the FSI
appears not to  be based on higher  order terms in the  formally known IA
series and alternative steps have been suggested instead.

Although demonstrated only  for the GRS theory $^{24}$ (cf. (Eq. 3.17b)),
it has first been assumed that also in the IA the TI response or TICS may
be written as a  convolution in either $\omega$ or $y$ of  the PWIA and a
FSI  factor $R$.   The  latter  has subsequently  been  postulated to  be
related to  the optical potential,  which by its very  definition affects
only the knocked-out proton.  By hindsight that proton is then correlated
to   target   nucleons   in   a   manner   which   has   elsewhere   been
criticized$^{28}$.  The ultimate result actually resembles the IA version
of the first  cumulant approximation, discussed in Section  3c and judged
there to  lack a  theoretical basis.   In the  end the  actual difference
between the result in Ref. 14 and  Eq. (3.29) is the replacement there of
the momentum distribution, typical for the GRS, by a spectral function.

The IA predictions based on the above rules fit the data very  well, in
particular the  high energy  loss region.  The non-negligible large-$E$
strength of  the NM  spectral function  may well  be responsible  for the
agreement with the  large-$\omega$ data.  However, also  Benhar $et\, al$
get too  low cross sections  in the  low $\omega$ region  by conventional
means. Colour transparency has been called in there$^{36}$, and as a true
$deus\,ex\,machina$ it brings about nearly perfect agreement.

We leave for the moment the outcome of the $(e,e'p)$ NE18 SLAC experiment
which for $Q^2$, relevant for the data discussed here as well as for much
higher  values, does  not evidence  the presence  of colour  transparency
$^{37}$.  We rather address more conventional considerations,
and mention in particular
uncertainties inherent  in the  first cumulant  approximation (or  to any
other current description).  We mention  thus far disregarded effects and
corrections which the low $\omega$ region may well be sensitive to.

i)  It has  been assumed  throughout, that  FSI are  dominated by  binary
collisions described  in the  first cumulant approximation.   Ternary and
higher order collisions, though small in general may in particular affect
the sensitive low intensity, low $\omega$ area.

ii)  Even assuming  dominance of  binary collisions,  the parametrization
used  for the  underlying  $NN$ scattering  amplitude neglects  off-shell
effects, which in  a  NR  model we  found to  be
appreciable$^{10}$.   Benhar   $et\,  al$  moreover  used   the  simplest
parametrization, neglecting finite  range and the real part  of the above
amplitude.  Their effect is appreciable.

iii)  Benhar  $et\, al$  assumed  the  2-particle density  matrix  (3.10)
governing the binary collision contributions  to be diagonal, i.e. simply
replaced $\zeta_2$  in (3.11) by the  diagonal pair-distribution function
$g(r)$.   The  authors of  Ref.  14  themselves noticed  sensitivity  for
changes in diagonal correlations and the  same  holds   for  the
non-diagonal $\zeta_2$ in the approximation (3.11).

The above points have not been  raised in order to show deficiencies, but
serve instead  as a warning against  a search for perfect  agreement: the
approach  of Benhar  $et\,al$  as well  as  of any  other  one today,  is
inflicted  by  a number  of  intrinsic  uncertainties.  Those  should  be
compounded to  inaccuracies in the  method and accuracy of  extraction of
the quasi-data.

We only briefly mention work by Frankfurt $et\,al\,^{17}$  who address by
means of light cone dynamics ratios of TICS for finite nuclei to which we
hope  to  return  elsewhere. The same holds for a relativistic
Hamiltonian approach $^{38}$ with  attempted applications on TICS$^{39}$.

This  concludes  our  description  of inclusive  electron  scattering  of
several  GeV  from nuclear  matter.   Although  there  is room  for  even
substantial  improvements,  we  hope  to  have  shown  that  a  realistic
description  of selected,  relatively high-energy scattering  on nuclear
targets is presently possible and that unavoidable approximations are
largely under control.

\par

{\bf Acknowledgements}
\par
\noindent
The authors thank Donal Day, Adelchi Fabrocini and  Omar Benhar for
having
provided tabulated data and the non-nucleonic  parts of the $eN$ ICS.

%\noindent
%\vfil
%\eject

{\bf References}
\bigskip

\refn{$^1$}
D.B. Day $et\, al$, Phys. Rev. Lett. {\bf 43} (1979) 1143;
$ibid.\,${\bf 59} (1987) 427.

\refn{$^2$}
B.D. Serot and J.D. Walecka, Adv. Nucl. Phys. {\bf 16} (1986) 1.

\refn{$^3$}
S.A.   Gurvitz, Phys.   Rev. {\bf  C42}  (1990) 2653;  L.S.  Celenza,  W.
Koepf and C.M Shakin, Phys.  Rec  {\bf C42} (1990) 1989; W.  Koepff, L.S.
Celenza and  C.M.  Shakin  , Phys.   Rev. {\bf  C43} (1991) 425; Phys.
Rev. {\bf C44} (1991) 2130.

\refn{$^4$}
D.B. Day $et\, al$, Phys. Rev. {\bf C40} (1989) 1011.

\refn{$^{5}$}
H.A.  Gersch,  L.J.  Rodriguez and Phil  N.  Smith, Phys.  Rev.  {\bf A5}
(1972) 1547.

\refn{$^{6}$}
M.L.   Ristig and  J.W.  Clark,  Phys.  Rev.  {\bf B40}  (1989) 4355;  D.
Ceperley,  Momentum Distributions,  Eds.  R.N.   Silver and  P.E.  Sokol,
Plenum  Press  1989,  and  private communication;  C.   Carraro,  private
communication.

\refn{$^{7}$}
R.N. Silver, Phys. Rev. {\bf B37} (1988) 3794; $ibid. \,${\bf B38}
(1989) 2283.

\refn{$^{8}$}
A.S. Rinat, Phys. Rev. {\bf B40} (1989) 6625; $ibid. \,$ {\bf B42}
(1990) 9944; A.S. Rinat and M.F. Taragin, Phys. Rev. {\bf B41}(1990)
4247; E. Pace, , G. Salm$\grave e$ and A.S. Rinat, submitted to
Nucl. Phys. ${\bf A}$.

\refn{$^{9}$}
C. Carraro and S.E. Koonin, Phys. Rev. Letters {\bf 65} (1990) 2792.

\refn{$^{10}$}
A.S. Rinat and M. Taragin, Phys. Lett. {\bf B267} (1991) 447.

\refn{$^{11}$}
M. Levi and J. Sucher, Phys. Rev. {\bf 186} (1969) 1656.

\refn{$^{12}$}
M.N. Butler and S.E. Koonin, Phys. Lett. {\bf B205} (1988) 123.

\refn{$^{13}$}
e.g. C. Ciofi degli Atti, S. Liuti and S. Simula,
Phys. Rev. {\bf C41} (1990) R2472

\refn{$^{14}$}
O. Benhar, A. Fabrocini, S. Fantoni, G.A. Miller, V.R. Pandharipande
and I. Sick, Phys. Rev. {\bf C44} (1991) 2328; $ibid$ {\bf D44} (1991)
692.

\refn{$^{15}$}
S. Fantoni and V.R. Pandharipande, Nucl. Phys. {\bf A473} (1987) 234;
A. Fabrocini and S. Fantoni, Nucl. Phys. {\bf A503} (1989) 375.

\refn{$^{16}$}
O. Benhar, A. Fabrocini and S. Fantoni, Nucl. Phys.
{\bf A505} (1989) 267;
A. Ramos, A.Polls and W.H. Dickhoff, Nucl. Phys. {\bf A503} (1989) 1,
and to be publ. B.E. Vonderfecht, Washington Univ. St. Louis PhD. Thesis,
May '91.

\refn{$^{17}$}
L.L. Frankfurt, M.I. Strikman, D.B. Day and M. Sargsyan, INPP preprint
1992.

\refn{$^{18}$}
e.g. P.J. Mulders, Physics Reports {\bf 185} (1990) 83.

\refn{$^{19}$}
e.g. H. Meier-Hadjuk, U. Oelfke and P.U. Sauer, Nucl. Phys.
{\bf A 499} (1989) 637; K. Nakano, Nucl. Phys {\bf A511} (1990) 664;
P.J. Mulders, A.W. Schreiber and H. Meyer, Nucl. Phys. {\rm bf A549}
(1992) 498.

\refn{$^{20}$}
A. Bodek and J.L. Ritchie, Phys Rev. {\bf D23} (1981) 1070.

\refn{$^{21}$}
A.E.L. Dieperinck, T. De Forest Jr., I. Sick and R.A. Brandenburg,
Phys. Lett. {\bf B63} (1976) 261.

\refn{$^{22}$}
T. de Forest, Jr., Nucl. Phys. {\bf A392} (1983) 232.

\refn{$^{23}$}
G.B. West, Phys. Reports {\bf C18} (1975) 264.

\refn{$^{24}$}
H.A. Gersch and L.J. Rodriguez, Phys. Rev. {\bf A8} (1973) 905.

\refn{$^{25}$}
C. Carraro and A.S. Rinat, Phys. Rev. {\bf B45} (1992) 2945.

\refn{$^{26}$}
S.A. Gurvitz, Phys. Rev. {\bf B42} (1990) 2653.

\refn{$^{27}$}
A. Kohama, K. Yazaki and R. Seki, Nucl. Phys. {\bf A536} (1992) 716;
$ibid$ {\bf A551} (1993) 687.

\refn{$^{28}$}
A.S. Rinat and B.K. Jennings, submitted to Nucl. Phys.

\refn{$^{29}$}
e.g C. Ciofi degli Atti, E. Pace and G. Salm$\grave e$,
Phys. Rev. {\bf C43} (1991) 1155.

\refn{$^{30}$}
A.S. Rinat and W.H. Dickhoff, Phys. Rev. {\bf B42} (1990) 10004.

\refn{$^{31}$}
e.g. A.S. Rinat and R. Rosenfelder. Phys. Lett. {\bf B193}
(1987)
411; N. Poliatzki and S.A. Gurvitz, Nucl. Phys. {\bf A524} (1991) 217.

\refn{$^{32}$}
I.E. Lagaris and V. R. Pandharipande, Nucl. Phys. {\bf A359} (1981) 331.

\refn{$^{33}$}
S. Galster $et\,al$ Nucl. Phys {\bf B32} (1971) 221.

\refn{$^{34}$} J. Bystricki, F. Lehar and Z. Janout, Note CEA-N 1547(E);
W. Grein, Nucl. Phys. {\bf B131} (1977) 255; A.V. Dobrovolski $et\,al$,
Nucl. Phys. {\bf B214} (1983) 1; B.H. Silverman $et\,al$, Nucl. Phys.
{\bf A499} (1989) 763.

\refn{$^{35}$}
O. Benhar, private communication.

\refn{$^{36}$}
e.g. G.R. Farrar, H. Liu, L.L. Frankfort and M.I. Strikman, Phys. rev.
Lett. {\bf 61} (1988) 686.

\refn{$^{37}$}
B. Fillipone, Invited talk, PANIC XIII, Perugia, Italy, 1993.

\refn{$^{38}$}
e.g. instance F. Coester in  Chap. 5, Proc. Int'l Spring School on
Medium and High-Energy Nuclear Physics, Taipeh Taiwan, 1988.

\refn{$^{39}$}
B. Keister and M.N. Butler, private comm.
\endpage
%\vfil
%\eject

{\bf  Figure captions}

\par\offset{55pt}{Fig. 1.} Diagram representing the lowest order
contribution in a PWIA series for the inclusive
scattering amplitude of a virtual
photon from a target.

\par\offset{55pt}{Fig. 2a} Quasi-data for TICS for NM from Ref. 1
for electron scattering energy and angle $\epsilon_{el}=2.02$ GeV,
$\theta=15^{\circ}$. Prediction are from Eq. (3...),
the first cumulant approximation  in the GRS theory. Dashes are
asymptotic  results, drawn line includes FSI. Data have been cut on
higher $\omega$ side by lack of reliable estimates
for inelastic processes.

\par\offset{55pt}{Fig. 2b} Same as Fig. 2a for $\theta=20^{\circ}$.
++ give results for the zero-range test.

\par\offset{55pt}{Fig. 2c} Same as Fig. 2a for $\epsilon_{el}$=3.595 GeV;
$\theta=16^{\circ}$.

\par\offset{55pt}{Fig. 2d} Same as Fig. 2c for $\theta=20^{\circ}$.
Inelastic contributions have been included and full
$\omega$ range has been retained. Second dashed curves is asymptotic
elastic and inelastic TICS.

\par\offset{55pt}{Fig. 2e} Same as Fig. 2d for $\theta=25^{\circ}$.

\par\offset{55pt}{Fig. 2f} Same as Fig. 2d for $\theta=30^{\circ}$.

\par\offset{55pt}{Fig. 2g} Same as Fig. 2b for $\epsilon_{el}$=3.995 GeV,
$\theta=30^{\circ}$.

\par\offset{55pt}{Fig. 3a} Same as Fig. 2a. Prediction are first
cumulant approximation based on
two lowest terms in the $1/q$ expansion for the IA theory. Dot-dashes,
dottes and dashes correspond to avarage separation energies $\lbrace
\Delta\rbrace$=0.0, 30 and 60 MeV. For comparison we also entered the
GRS result of Fig. 2a by a drawn line.
Data have been cut on higher $\omega$ side for lack of reliable estimates
of inelastic contributions.

\par\offset{55pt}{Fig. 3b} Same as Fig. 3a for $\theta=20^{\circ}$.
++ give results for the zero-range test.

\par\offset{55pt}{Fig. 3c} Same as Fig. 3a for $\epsilon_{el}$=3.595 GeV;
$\theta=16^{\circ}$.

\par\offset{55pt}{Fig. 3d} Same as Fig. 3c for $\theta=20^{\circ}$. Full
$\omega$ range has been retained.

\par\offset{55pt}{Fig. 3e} Same as Fig. 3d for $\theta=25^{\circ}$.

\par\offset{55pt}{Fig. 3f} Same as Fig. 3d for $\theta=30^{\circ}$.

\par\offset{55pt}{Fig. 3g} Same as Fig. 3b for $\epsilon_{el}$=3.995 GeV,
$\theta=30^{\circ}$.

\par\offset{55pt}{Fig. 3g} Same as Fig. 3d for $\epsilon_{el}$=3.995 GeV,
$\theta=30^{\circ}$.

\par\offset{55pt}{Fig. 4} Data on the scaling function $G^{el}(q,y_W^r)$,
Eq. (7.2a), as function of the relativistic West scaling variable
(3.8$'$). $'$Inelastic$'$ parts have been removed.

\par\offset{55pt}{Fig. 5} Same as Fig. 4 for the scaling function
$G^{el}(q,y_0^r)$, Eq. (7.2b), as function of the relativistic IA
scaling variable (3.26$'$).

\endpage

\ctrline {\bf Table I}
\medskip
\begintable
$q$ & $T_q$ | $\sigma^{tot}_{pp}$ & $\sigma^{reac}_{pp}$|
$\sigma^{tot}_{pn}$ & $\sigma^{reac}_{pn}$ | $\tau_{pp}$ & $\tau_{pn}$|
$(Q_0)_{pp}$ & $(Q_0)_{pn}$ \nr
(GeV) & (GeV) |(mb) &(mb) | (mb) & (mb)|
 (GeV) & (GeV)\crthick
 0.5~ & 0.125 | 25.8 & 0 | 55 & 0 | 1.0 & 1.0|
$\sim$1.83 & 1.83 \nr
0.75 & 0.262 | 24.3 & 0 | 38 & 0 | $\sim$0.7 &
$\sim$0 | 1.12 & 1.12 \nr
1.00 & 0.432 | 28.1 & 4.0 |37 & 5 | 0.5 & -0.18|
  0.91 & 0.91 \nr
  1.20 & 0.585 | 36.6 & 12.6| 36 & $\ge$8 | 0.4 & -0.25|
  0.84 & 0.84 \nr
  1.31 & 0.673 | 42 & 15.7 | 35.8 & $\ge$10 | 0.3 &
-0.273 | 1.34 & 1.45 \nr
1.40 & 0.747 | 46.9 & $\sim$17 | 39.4 &
$\ge$18.4 | $\sim$0 & $\sim$-0.35 | 0.68 & 0.67 \nr
1.50 & 0.831 | 47.8
&22.8 | 38. & $\sim$20 | -0.1 & $\sim$-0.38 | 0.65 & 0.65 \nr
1.60 &
0.916 | 47.5 & 22.5 | 40 & $\sim$ 23 | -0.14 & -0.41 | 0.63 & 0.63 \nr
1.75 & 1.047 | 46 & 22 | 41 & $\sim$ 23 | -0.25 & -0.40 | 0.60 & 0.60 \nr
2.00 & 1.271 | 47.0 & 25 | 42 & 22 | -0.22 & -0.40 | 0.59 & 0.59 \nr
2.20 & 1.453 | 47.0 & 26 | 42 & 28 | -0.18 & -0.41 | 0.59 & 0.59 \nr
2.40 & 1.638 | 46 & 26 | 42 & 28 | -0.15 & -0.40 | 0.59 & 0.59
 \endtable
\medskip
\ctrline{Partly interpolated  parameters for elastic $pp$ and $pn$
scattering$^{34}$}

\end